\documentclass[doublecol]{epl2} 

\usepackage{verbatim}
\usepackage{amsmath}
\usepackage{enumitem}

\newcommand{\sg}{spin glass}
\newcommand{\sgs}{spin glasses}

\newcommand{\ham}{\ensuremath{H}}

\newcommand{\zerop}{zero-particle}
\newcommand{\onep}{one-particle}

\newcommand{\twop}{two-particle}

\newcommand{\threep}{three-particle}

\newcommand{\ipr}{\text{IPR}}

\newcommand{\beq}{\begin{equation}}
\newcommand{\eeq}{\end{equation}}
\newcommand{\kt}{\rangle}
\newcommand{\br}{\langle}
\newcommand{\f}{\frac}
\newcommand{\rmd}{\ensuremath{\mathrm{d}}}

\newcommand{\ket}[1]
{
\ensuremath{\left|#1 \right\rangle}
}
\newcommand{\avg}[1]
{
\ensuremath{\left\langle #1 \right\rangle}
}

\title{Persistent entanglement in a class of eigenstates of quantum Heisenberg \sgs}
\shorttitle{Persistent entanglement in quantum \sgs} %Insert here a short version of the title if it exceeds 70 characters

\author{Arun Kannawadi\inst{1} \and Auditya Sharma\inst{2} \and Arul Lakshminarayan\inst{3}}
\shortauthor{Kannawadi \etal}

\institute{                    
  \inst{1} Department of Physics, Carnegie Mellon University, Pittsburgh, Pennsylvania, USA\\
  \inst{2} Department of Physics, Indian Institute of Science Education and Research, Bhopal, India\\
  \inst{3} Department of Physics, Indian Institute of Technology Madras, Chennai, India.
}

\abstract{ The eigenstates of a quantum \sg\ Hamiltonian with long-range interaction are examined from the point of view of localisation and entanglement. 
In particular, low particle sectors are examined and an anomalous family of eigenstates is found that is more delocalised but also has larger inter-spin entanglement. 
These are then identified as particle-added eigenstates from the \onep\ sector. 
This motivates the introduction and the study of random promoted \twop\ states, and it is shown that they may have large delocalisation such as generic random states and scale exactly like them.
However, the entanglement as measured by two-spin concurrence displays different scaling with the total number of spins. This shows how for different classes of complex quantum states entanglement can be qualitatively different even if localisation measures such as participation ratio are not. }

\begin{document}

\maketitle

\section{Introduction}

Consider a Hamiltonian of $L$ spin-$1/2$ particles that conserves total spin in some direction; for definiteness, let  $\sigma^z_T=\sum_i \sigma^z_i$ be conserved. 
The Hamiltonian is rendered block-diagonal in the $\sigma^z$ basis and the blocks are specifed by the total spin $\sigma^z_T$, the most trivial of them being states where all the spins are up or down.
The case when $m$ of the spins are up (corresponding to $\sigma^z_T=(L-2m)/2$) is a $\binom{L}{m}$ dimensional subspace.
For example, $m=1$ states are used to transport information across spin-chains~\cite{Sougato07}. We will refer to states with $m$ up spins as ``$m-$ particle states", since similar block-diagonal structure in the Hamiltonian appear in spinless fermion models.

%\begin{comment}
%Since they are the simplest subspaces, we also
%concentrate on them, with a particular focus on single and
%two-particle (or magnon) sectors, which may not contain the ground
%state.  
%\end{comment}
In \onep\ states, there is a clear monotonic
relationship~\cite{ArulSub, EntLoc1, EntLoc2} between localisation, for
example as measured by the participation ratio (for e.g., see
~\cite{PhysRevE.82.031130, PhysRevE.81.036206, PhysRevA.82.011604,
  PhysRevE.87.012125}), and the inter-spin entanglement as measured by,
say, concurrence \cite{Wooters,Woottersentform}: more the localisation, less the
entanglement.  
There is no such strict monotonic relationship between localisation
and entanglement for states with higher particle-number.  However, there are statistically very significant
correlations between localisation and entanglement~\cite{Karthik07,Brown:08,dukesz2009interplay,beugeling:2015,viola2007generalized,giraud2007entanglement}. 
%Also cite the following:
%
%. Giraud, J. Martin, and B. Georgeot,
%Entanglement of
%localized states
%, Phys. Rev. A
%76
%, 042333 (2007).
%[38]
%  L.  Viola  and  W.  G.  Brown,
%Generalized  entanglement
%as a framework for complex quantum systems:  Purity vs
%delocalization measures
%, J. Phys. A
%40
%, 8109 (2007).
It is shown below that, for \twop\ states, in contrast to \onep\ states, \emph{on average} (in a way to be defined precisely later), increased localisation implies {\it enhanced} two-spin entanglement as measured by concurrence. This effect is even more pronounced for \twop\
eigenstates of a \sg\ Hamiltonian studied below. It should be emphasised that this is entanglement between two spins - other  measures such as block entropy may well decrease
with localisation.

To study this in the simplest statistical context, {\it random} states of definite particle-number were considered using an ensemble that
was uniformly distributed in such subspaces~\cite{Vikram11}.  It was found that while the expected entanglement
between two spins for \onep\ states having $L$ spins scales as $1/L$ and that of \twop\ states scale as $1/L^2$, in the case of three or more particle states, entanglement is practically absent and is 
exponentially small in $L$ ($\exp(-L \ln L)$, to be precise).  This is consistent with such states
having larger multipartite entanglement and the fact that entanglement
moves away from being locally shared.  In some sense, the
``environment" of any two spins is too large for the entanglement between the spins to remain intact. 

This scenario is observed in models of many-body localisation, for example, the XXZ model with a random external field~\cite{Santos2004,Carlos2005,Bera2016}.
In this case, when the interaction dominates, disordered eigenstates in the half-filled ($m=L/2, \, \sigma^z_T=0$) sector are such that there is vanishing concurrence between two spins. Along with a many-body localisation transition, concurrence also arises to once again slowly disappear when the disorder completely dominates the interaction. The present work must therefore be seen in the larger context of entanglement in disordered interacting quantum systems.

In random states, it is \emph{almost impossible} to find
entanglement between subsystems unless the block length (size of the
subsystem) is of the order of the size of the (pure) system, typically
$\sim L/2$~\cite{kendon2002bounds,kendon2002typical,scott2003entangling, Bhosale2012}. 
However, when the states are restricted to be in the subspace of a fixed particle number, then one \emph{can} find entanglement as long as the particle-number does not exceed the block length~\cite{Vikram11}. 
To reiterate, for example, concurrence between two spins can be found in a typical \twop\ state but not in typical or random \threep\ state.

The present work identifies a subset from within the subspace of definite-particle states that have enhanced entanglement. These are simply particle-added states from lower particle-number sectors referred here as ``promoted states". One can generate a whole class of \emph{random} promoted states, with high entanglement (compared to generic random definite-particle states), and a different scaling with the total number of spins $L$.

In order to compare these statistical considerations with physical systems,
we study the eigenstates of  the infinite-range quantum Heisenberg \sg. The isotropy of the Hamiltonian implies that total spin along {\it any} direction is conserved, which in turn allows promoted eigenstates to exist (see later). 
A subset of \twop\ eigenstates is found to have pronounced entanglement. A closer scrutiny shows that these are in fact obtained by promoting \onep\ eigenstates. 
This adds a new dimension to the study of entanglement in  many-body systems \cite{Amico08} with spin rotational symmetry. One can expect to find the analysis in this paper to be relevant for generic non-integrable Hamiltonians that have such a symmetry. 

%The structure of the paper is as follows. In
%Sec.~\ref{sec:formulation}, we describe the \sg\ Hamiltonian to
%motivate the problem and the quantum properties of the system we are
%interested in. We also highlight the existence of promoted
%eigenstates, and how they are related to a symmetry. In
%Sec.~\ref{sec:promoted}, we obtain a few analytical results for the
%quantities described in Sec.~\ref{sec:formulation} when the system is
%in a `random promoted' state subject to certain assumptions. 
%We then compare these results with two specific \sg\ models namely the
%`infinite-range' and `nearest-neighbour' models.  
%In Sec.~\ref{sec:powerlawdecay}, we show systematically that the
%`promoted' eigenstates, that were distinguishable from the rest of the
%eigenstates from an entanglement perspective in the case of the
%infinite-range \sg\ model, become indistinguishable as the range of
%interaction between the spins becomes smaller.
%The last section summarizes the conclusions of our work, and offers an outlook for
%future work.

\section{Formulation of the problem}
\label{sec:formulation}
\subsection{The infinite-range quantum Heisenberg \sg}
The Hamiltonian considered in this paper is 
\begin{align}
\label{eq:ham}
\ham =  \sum_{\substack{{j=1}\\{i > j}}}^{L} J_{ij} \vec{\sigma_i}.\vec{\sigma_j}
\end{align}
where $J_{ij}$ are independent random variables drawn from the normal distibution $\mathcal{N}(0,1)$. This work concentrates on the eigenstates and therefore the normalisation of the energy is irrelevant.
The Hamiltonian takes a block diagonal form, where each block is
characterised by a particle-number $m$, which is also the total number of up-spins in the $z$ direction.

The $m$-particle basis states are $\ket{i_m\dots i_1}$ where $i_1<i_2 <\dots< i_m$ refer to positions of  `up-spins':$\ket{\uparrow}$ in the $\sigma_z-$ basis, the others being down.  For a fixed $m$, the state with uniform superposition of all the basis states is necessarily an eigenstate, with eigenvalue
$S_J = \sum_{i>j} J_{ij}$.  This is a direct consequence of the isotropy of the Hamiltonian which
implies $[\mathcal{H},\sigma^{\pm}_T] = 0$, where $\sigma^\pm_T = \sum_i\sigma_i^\pm = \sum_i \left(\sigma_i^x \pm \sigma_i^y\right)$. This in turn implies that repeated
action of the $\sigma^+_T$ operator on the zero-particle eigenstate $\ket{\downarrow}^{\otimes L}$ will also give eigenstates. We will
refer to such eigenstates as `all-one' states of the appropriate particle-number.
\begin{equation}
  \ket{\text{all-one state}}_{m} \propto \left(\sigma^+_T\right)^m \ket{\downarrow}^{\otimes L}  \propto  \sum_{i_1<\dots<i_m} \ket{i_m\dots i_1}.
 \label{eq:allone}
\end{equation}

\subsection{Measures of Entanglement \& Localisation}
\label{subsec:measures}
\begin{figure*}
  (a)\includegraphics[width=0.9\columnwidth]{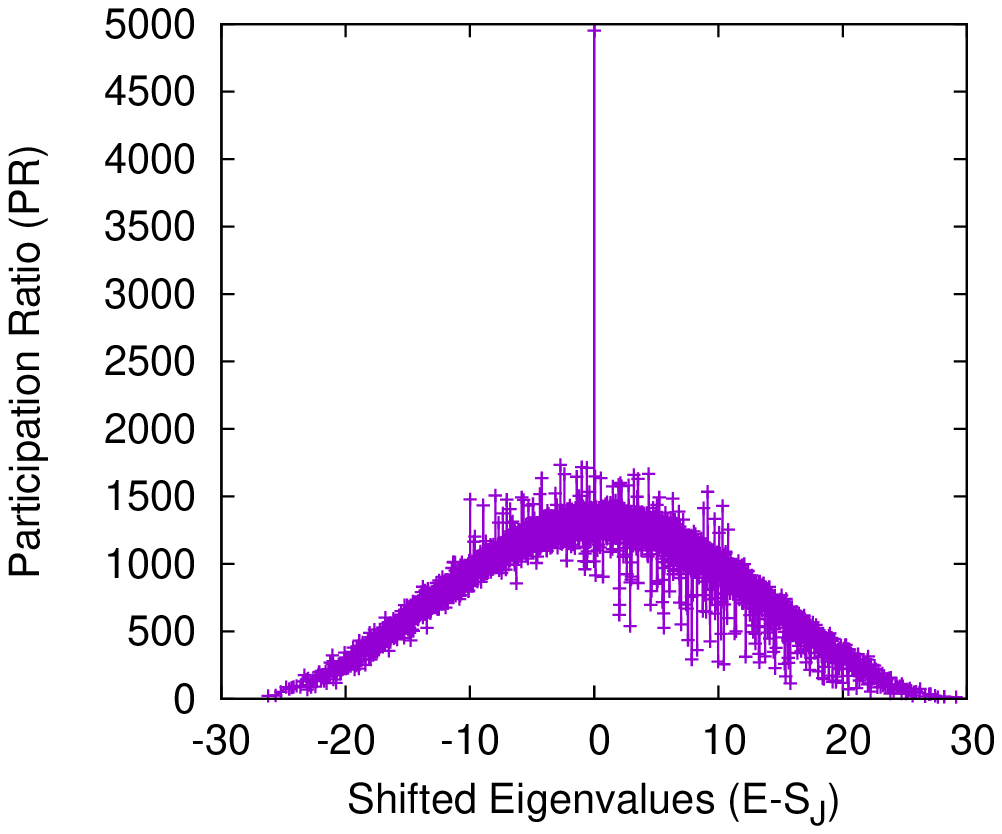}\ 
  (b)\includegraphics[width=0.9\columnwidth]{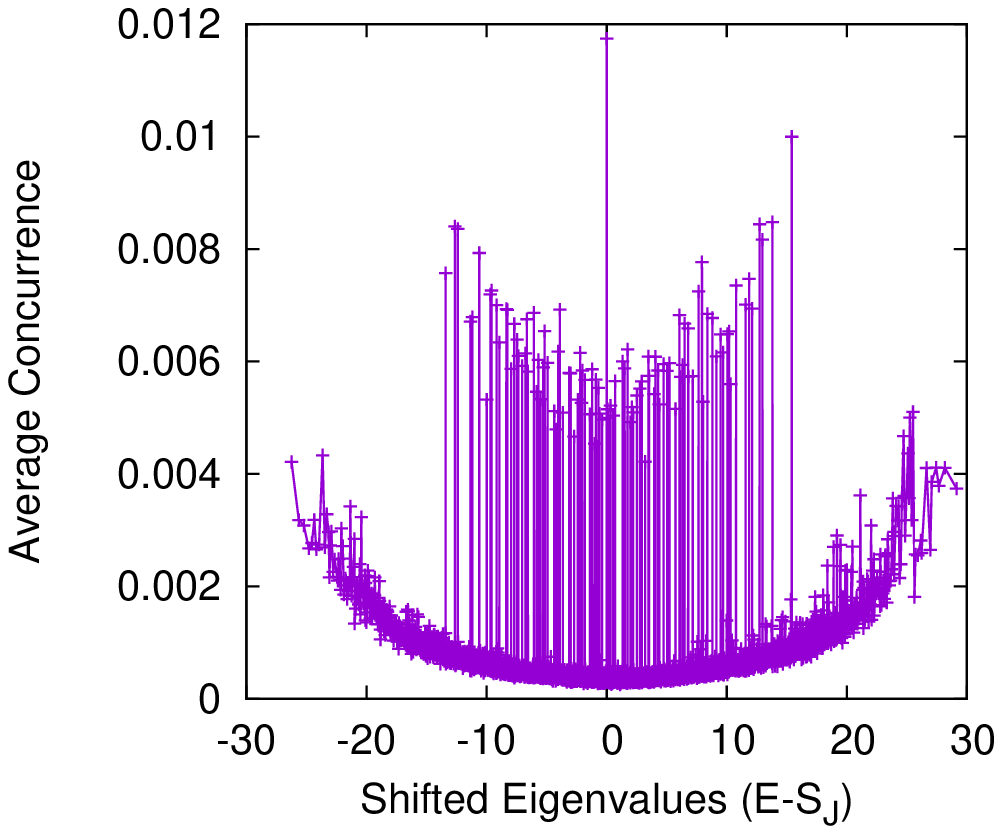}
  \caption{Plot of the participation ratio and average
    concurrence (averaged over all pairs of spins) of all the eigenstates, arranged by their eigenvalues, for a typical realisation of the SK \sg\ model with $L=100$ and $m=2$.
    $S_J = \Sigma_{i>j}J_{ij}$ is the energy of the all-one eigenstate. The
    broad features are similar for \emph{any random}
    realisation of the SK Hamiltonian.}
  \label{fig0}
\end{figure*}
This work focuses on bipartite entanglement between two spins as measured by concurrence \cite{Wooters}.
Concurrence is a simply calculable entanglement monotone within any two-level system which can be in a mixed or pure state. For a two spin state $\rho$, $C(\rho) \equiv \text{max}(0,\lambda_1-\lambda_2-\lambda_3-\lambda_4)$,
in which $\lambda_1 \ge \lambda_2 \ge \lambda_2 \ge \lambda_4$ are the eigenvalues of the positive matrix $R = \sqrt{\sqrt{\rho}\tilde{\rho}\sqrt{\rho}}$ with $\tilde{\rho} = (\sigma^y\otimes\sigma^y)\rho^{*}(\sigma^y\otimes\sigma^y)$.
It is known that $0 \le C(\rho) \le 1$ and it is $0$ iff $\rho$ is a separable state and is $1$ iff the state is maximally entangled. For definite-particle states, it is known that the reduced density matrix of any
two spins takes the following form~\cite{Connor2001}:
\begin{align}
\rho = 
\begin{pmatrix}
v & 0     & 0 & 0\\
0 & w     & z & 0\\
0 & z^{*} & x & 0\\
0 & 0     & 0 & y
\end{pmatrix}.
\label{eq:rho}
\end{align}
In this case, the concurrence is simply given by ~\cite{Connor2001}:
\begin{align}
\label{eq:concurrence}
C(\rho) = \max\left({2(|z|-\sqrt{vy}),0}\right). 
\end{align}
The quantity $2(|z|-\sqrt{vy})$ is often referred to as \emph{pre-concurrence}.
Throughout this work, we will refer to ``average concurrence" of a
state as the average of the concurrence values between all pairs of spins. %and `expectation value of
%concurrence' of a \emph{class of states} as the double average of the
%concurrence values between all pairs of spins and over all states that
%belong to that particular class.

The inverse participation ratio (see for e.g., ~\cite{PhysRevE.82.031130, PhysRevE.81.036206,
  PhysRevA.82.011604, PhysRevE.87.012125}) is a basis-dependent
quantity that quantifies localisation and is defined as follows. If a state $|\psi\rangle =
\sum_{\beta}a_{\beta}\ket{\beta}$, where $\ket{\beta}$ are the kets in the
computational basis, then $\ipr \equiv \sum_{\beta} a_{\beta}^{4},$ from which the
participation ratio $\text{PR} \equiv 1/\ipr$ is
obtained.  The range of values that the participation ratio of any
state can assume is $\left[ 1, D\right]$, where $D$ is the
dimensionality of the Hilbert space.  $PR=1$ occurs when the given
state happens to be one of the basis vectors themselves, with all but
one of the coefficients being zero and is thus highly localised.
$PR=D$ occurs when the magnitude of each of the coefficients is the
same and therefore corresponds to a highly delocalised state. Thus, for the
all-one state given by Eq.~\ref{eq:allone}, $PR=D=\binom{L}{m}$ is the largest possible. 

\section{Concurrence and PR in the eigenstates}
%The infinite-range Heisenberg \sg\ model (SK model) is given by: $J_{ij} = \mathcal{N}(0,1),$ for $ \forall \, i>j$. Here
%$\mathcal{N}(\mu,\sigma^2)$ stands for the normal distribution with
%mean $\mu$ and variance $\sigma^2$. %We impose periodic boundary conditions, so that $j=0$ corresponds to $j=L$. 
%We are concerned here only with the properties of the eigenstates, therefore the normalization of the interaction matrix does not matter.
 Fig.~\ref{fig0} shows the participation ratio and the average
 concurrence of all the eigenstates of the $m=2$ sector for one
 realisation of the Hamiltonian in Eq.~\ref{eq:ham} with $L=100$. The central spike at $E-S_J = 0$ in each of Fig.~\ref{fig0}(a) and Fig.~\ref{fig0}(b)
 corresponds to the \twop\ all-one eigenstate. While no other
 significant spikes are found in Fig.~\ref{fig0}(a), several such
 spikes are found in Fig.~\ref{fig0}(b). 
%The eigenstates separate out into two distinct
% clouds in a phase diagram where the average concurrence is plotted
% against PR ($+$ points in Fig.~\ref{fig1}). 
It is now shown that these spikes in  Fig.~\ref{fig0}(b) are in fact the subset of two-particle eigenstates obtained by promoting \onep\ eigenstates. Closer scrutiny shows that the small structures present in the particpation ratio do not correspond, by and large, to the well delineated ones in the average concurrence figure, and hence these states
with enhanced entanglement are not special as far as localisation  is concerned.

\subsection{Symmetries of the Heisenberg Hamiltonian}
The \emph{particle-number operator} $\hat{N}_{\uparrow} \equiv \sum_k \frac{\sigma_k^z + 1}{2}$ has eigenstates which are definite-particle states, with 
corresponding particle-numbers as their eigenvalues.
The definite-particle nature of the Hamiltonian is a result of $H$ commuting with $\sigma^z_T = \sum_{i=1}^{L} \sigma_i^z$ and hence with $\hat{N}_{\uparrow}$.
Due to the isotropy of the Hamiltonian, it also follows that the operators
$\sigma^{\pm}_T = \sum_{i=1}^{L} \sigma_{i}^{\pm}$ commute with $H$, where $\sigma_i^\pm = \sigma_i^x \pm i\sigma_i^y$.
This implies that if $|\psi\rangle$ is an eigenstate of $\ham$, then
$\sigma^{\pm}_T|\psi\rangle$ must also be an eigenstate of $\ham$ with
the same eigenvalue.  In addition, if $\ket{\psi}$ has particle-number
$m$, then $\sigma^\pm_T \ket{\psi}$ has particle-number $m \pm 1$.  The
particle-added state $\sigma^+_T |\psi\rangle$ is referred to as the
\emph{promoted $(m+1)$-particle state} corresponding to $\ket{\psi}$.

\subsection{Promoted States}
\label{sec:promoted}
%The \emph{particle-number operator} $\hat{N}_{\uparrow} \equiv \sum_k \frac{\sigma_k^z + 1}{2}$ acting on an \mparticle state $\ket{\psi}$ gives $ m\ket{\psi}$. 
%We define  $\sigma^+ = \sum_k \sigma_k^+$ as the
%\emph{promotion operator}.  
%%One can immediately identify this with
%%the ladder operator associated with the quantum harmonic oscillator.
%%AL: I removed this as I am not sure about its correctness and is not relevant.
%For an \mparticle state $\ket{\psi}$, $\hat{N}_{\uparrow}\sigma^+\ket{\psi} = (m+1)\left( \sigma^+\ket{\psi}\right)$, 
%which follows from the commutation relation $\commutator{\sigma_i^z}{\sigma_j^+} = 2\sigma_i^+\delta_{i,j}$.
%So $\sigma^+\ket{\psi}$ is an $(m+1)$-particle state, a state with its particle-number one higher
%than that of $\ket{\psi}$.  
%Is not all this a repetition??

\begin{figure}[ht]
\includegraphics[width=\columnwidth]{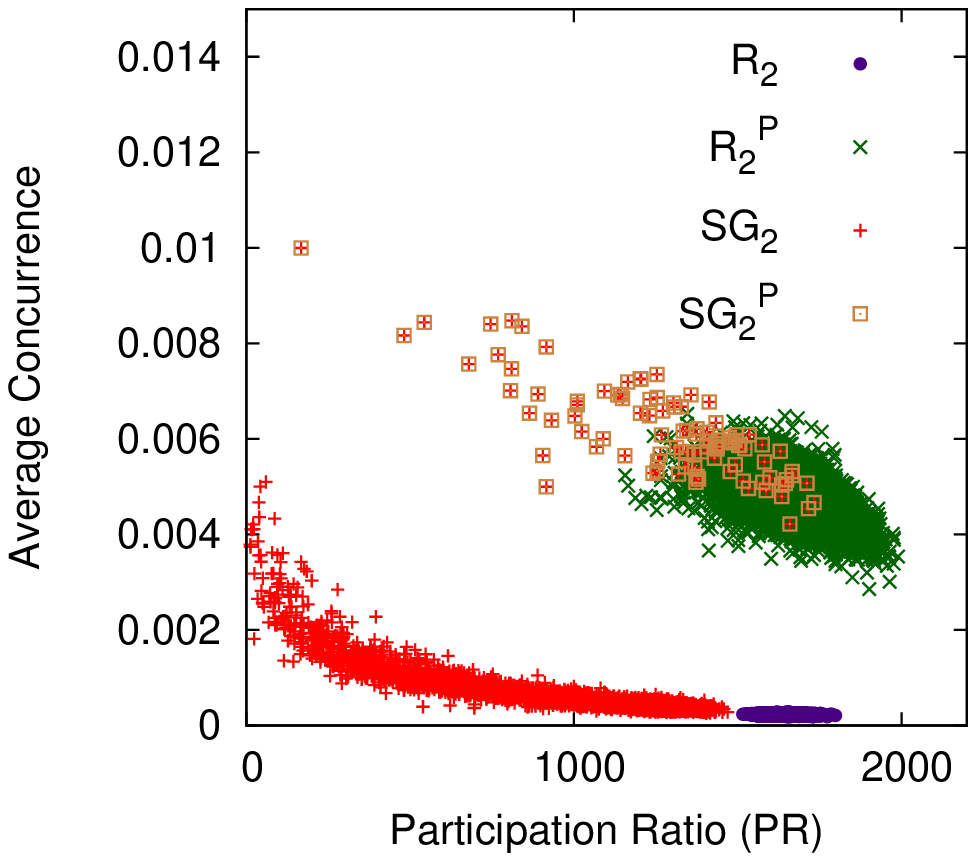}
\caption{Scatter plot of average concurrence vs. participation ratio
  (PR) of all the eigenstates in the $m=2$ sector ($SG_{2}$) of
  \emph{a typical} realisation of the infinite range quantum
  Heisenberg \sg\ for $L=100$. The promoted eigenstates
  ($SG_{2}^{P}$) are seen to form a separate cloud. Data for an equal
  number (4950) of \emph{random} $2$-particle states
  ($R_{2}$) and \emph{random-promoted} $2$-particle states
  ($R_{2}^{P}$) are included. The solitary `all-one' eigenstate with a participation ratio of $PR=4950$ is not shown.}
\label{fig1}
\end{figure}

Fig.~\ref{fig1} shows a plot of the average concurrence {\it vs}
participation ratio of all the eigenstates of the $m=2$ sector for
one realisation of the Hamiltonian in Eq.~(\ref{eq:ham}) with $L=100$. All the eigenstates are marked by a $+$ symbol. They are
seen to separate into two ``blobs", with the main group at the bottom
left, separated from a small set of eigenstates that have larger
concurrence and a majority of which are also more delocalised. The latter are shown to correspond to precisely the large spikes in  Fig.~\ref{fig0}(b). 

Promoted states are also eigenstates of the operator
$\sigma^+_T\sigma^-_T$ with a non-zero eigenvalue, a property we used to identify the promoted \twop\ eigenstates
amongst the full set of \twop\ eigenstates of the Hamiltonian. The
promoted eigenstates thus identified have a box enclosing the $+$ symbol
in Fig~\ref{fig1}. The promoted eigenstates indeed stand out and form a separate cloud with significantly higher average concurrence and predominantly higher delocalisation. 

Furthermore, we verified that the \threep\ eigenstates promoted from the \onep\ and \twop\ eigenstates also form distinct clouds when we consider $2$-spin and $3$-spin entanglement in the \threep\ sector (figure not included).  Thus, the promoted character of
these states is distinguished in these localisation-entanglement
plots. As a useful benchmark against which to compare, we also include data for random \twop\ states (circles in Fig.~\ref{fig1}), and
``random promoted" two-particle states (marked by $(\times)$ symbols
in Fig.~\ref{fig1}).  We defer a detailed definition and discussion of these states to the next section, but it is worth pointing out now that while the ``genuine'' random \twop\ states are different from ``genuine'' \twop\ eigenstates of the \sg\ Hamiltonian, the promoted \twop\ states in these two cases are not all that different. 
%This suggests that we might study the promoted states without worrying too much about their origir.

\section{Random promoted states} 
\label{subsec:random2p}
It is naturally of interest to understand the origin of the enhanced entanglement in the promoted states as compared to other states.
As a statistical model, states promoted from \emph{random} definite-particle states are
easier to study than the promoted eigenstates of random  Heisenberg Hamiltonians such as in Eq.~(\ref{eq:ham}).  
For the purposes of this work, it suffices to restrict to ensembles states with real coefficients, which are relevant to systems preserving time-reversal symmetry.%~\cite{ArulSub}.
%AL: removed citation to ArulSub as of course this is a much more basic statement and well-known before these authors.

The defining characteristic of the random states is that their distribution is isotropic in the associated Hilbert
space. To generate a random state, i.e., a state with random orientation in a Hilbert space of dimension $D$, we form a vector $\left( r_1, r_2, \dots r_D\right)$, where $r_k$ s are i.i.d. random variables drawn from a Gaussian distribution $\mathcal{N}(0, 1)$. 
%%Note that the variance of this does not mater, so I made it 1 (AL)
The coefficients of the normalised state are then obtained by dividing the vector by its norm. For example, see \cite{WoottersRandom}. Random definite particles states are those sampled uniformly from the associated definite particle sector of states.
%%Wootters Foundations of Physics http://link.springer.com/article/10.1007%2FBF01883491

The expectation value of concurrence is calculated as
\beq
\avg{C} =\dfrac{1}{\binom{L}{2}}\sum_{k>l} \avg{C_{k,l}},
\eeq
where the averaging for any individual pair
of spins comes from an ensemble average, either via a statistical
model such as in the case of random states or from an ensemble of
Hamiltonians, as in the \sg\ case.  For random \onep\ states, it is
known that $\avg{C} = 4/(\pi L)$ \cite{ArulSub}. Note that for
\onep\ states, $\avg{\ipr} = 3/L$, as the average IPR for any $D$
dimensional real random states is $3/D$ for large $D$ \cite{brody81}.

It follows that for a random \twop\ state $\ket{\psi} = \sum_{i>j}a_{ij}\ket{ij}$, its average IPR (
$\sum_{i>j}|a_{ij}|^4 $) is
\begin{equation}
\avg{\ipr} = \frac{3}{\binom{L}{2}} \approx \frac{6}{L^2}.
\label{eq:IPR_2p}
\end{equation}
To estimate $\avg{C}$, it is helpful to first compute the probability of a pair of spins being entangled, $P(C>0)$.  For the
ensemble of all real random \twop\ states, it has been shown~\cite{Vikram11} that $P(C>0) = 2\sqrt{2}/\sqrt{\pi L}$ and the expectation value of the concurrence is $16/(\pi^{3/2}L^{2})$.
%%We infer that the $2$-particle \emph{eigens}tates of this Hamiltonian tend to have a higher average concurrence (which can be by as much as $10$ times), and are also more localized than random \twop states. The distribution of points for the eigenstates of the infinite-range Hamiltonian is thus very different from what is expected for random states. This immediately suggests that entanglement properties of random states, such as scaling behavior \cite{Vikram11} may not be found for eigenstates of spin systems, at least those with two-body interactions.% ~\cite{Arun:2011}.
A procedure similar to the one in ~\cite{Vikram11} can indeed be used to calculate the expectation value of the \ipr\ and concurrence for {\it promoted} random states.

\subsection{Random Promoted $2$-particle States}
\label{subsec:randompromo2p}
Consider a random \onep\ state $\sum_{k}a_{k}\ket{k}$. The $a_{k}$\ s are obtained following the procedure outlined above and satisfy $\sum_k a_k^2=1$.
%by drawing $L$ independent identically distributed (i.i.d) random variables from a
%normal distribution $\mathcal{N}(0,1)$,
A promoted \twop\ state (unnormalised) is obtained by action of $\sigma^{+}_T$, on a
\onep\ state: 
\begin{equation}
\label{eq:promrand2}
\sigma^+_T \, \sum_k a_k \ket{k} = \sum_{i>j}
(a_{i}+a_j) \ket{ij}.
\end{equation}
Thus the special structure of promoted states is apparent here:
only $L$ random numbers determine the $\binom{L}{2}$ coefficients of a random promoted \twop\ state,
while a random \twop\ state, in general, depends on $\sim L^2/2$ independent random numbers.

To ease the theoretical calculations, it is convenient to substitute the sum of $L$ random numbers,
$\sum_{k} a_{k}$, by ($L$-times) its mean, which is $0$. It is a reasonable approximation since this
holds exactly for one-particle eigenstates of the above spin glass Hamiltonian (as a direct consequence of
orthogonality with respect to the all-one eigenstate).
%We checked numerically that the imposition or not of this constraint makes no practical difference.} 
It also follows that $\sum_{i>j}(a_i+a_j)=0$ if $\sum_k a_k = 0$.
The coefficients of the normalised promoted random \twop\ states are then
$a_{ij} = \left(a_i + a_j\right)/\sqrt{L-2}$, as $\sum_{i>j}(a_i+a_j)^2=L-2$.
%Without the additional approximation, the normalisation would depend on the
%coefficients themselves. 

%Since $a_k$\ s are random variables, $\sum_k a_{k}$ is also a random variable whose distribution is approximately $\mathcal{N}(0,1)$.
%Thus, even for large $L$, the deviation of $\sum_k a_{k}$ from its mean is significant.
%However, in the calculations that follow, we will replace $\sum_k a_k$ by its mean and assume for the sake of simplicity that 
%\begin{align}
%\sum_{k}a_{k} = 0, \label{eq:sum}
%\end{align}
%
%As we will show in Fig.~\ref{fig:entanglement_scaling}, this approximation is excellent.
%Note that this condition holds for the \onep eigenstates of the Hamiltonians of
%Eq.~\eqref{eq:generic_hamiltonian} (except for the `all-one' state) due to orthogonality with respect to the `all-one'
%eigenstate. 
%Comparison against numerical results for random \twop promoted states, without assuming Eq.~\ref{eq:sum}, shows that our this approximation is excellent.

The \ipr\ of a promoted \twop\ state can then be expressed in terms of the
\ipr\ of the corresponding \onep\ state:
\begin{equation}
 \label{eq:ipr_relation}
 \sum_{i>j}a_{ij}^4 = \frac{1}{(L-2)^2}\left((L-8) \sum_i a_i^4 + 3\right),
\end{equation}
where $a_{ij} = (a_i+a_j)/\sqrt{L-2}$, are the normalised
coefficients.  Note that Eq.~\ref{eq:ipr_relation} does not assume $L$
is large but only $\sum_k a_k = 0$. Thus, it implies that when $L<8$,
the IPR of the \twop\ state decreases with increase in the IPR of the
\onep\ state and vice versa and the trend changes for $L>8$.  We
observe in passing the somewhat amusing fact that when $L=8$, whatever may be the
\onep\ state, the promoted \twop\ state has an \ipr\ of
exactly $1/12$, again provided that the coefficients sum to zero.

Thus, as $\avg{\sum_k a_k^4} = 3/L$ it 
immediately follows that 
\beq
\avg{IPR} \sim \frac{6}{L^2}.
\eeq
Thus, to the leading order, this is {\it identical} to the IPR of 
``genuine'' \twop\ random states, as given in Eq.~\ref{eq:IPR_2p}.
Thus as far as localisation is concerned there is typically no difference between promoted and genuine \twop\ states, as also
confirmed by numerical data in Fig.~\ref{fig1}.  A very
different situation is found regarding quantum correlations, such as
entanglement, to which we now turn.

\subsection{Entanglement in promoted \twop\ states}
\label{subsec:entanglement_promo}
The elements of the two-spin reduced density matrix that are involved
in the entanglement between the two spins, as quantified by concurrence, are
$z$, $v$, $y$ (Eqs.~\ref{eq:rho},~\ref{eq:concurrence}).
For \twop\ states, when $\rho$ is the density matrix of spins at
positions 1 and 2 (which we consider for simplicity and without any
loss of generality), these elements are \beq y=a_{12}^2 ,\qquad z=
\sum_{k=3}^{L} a_{2k} a_{1k} , \qquad v=\sum_{\substack{k,l=3
    \\ k<l}}^{L}a_{kl}^2. \label{eq:vyz} \eeq Note that we are considering
\emph{real} state ensembles. 
%We may expect that while all the three quantities are random variables, $y$ being a single term has more
%fluctuation than the other two which are sums.
For generic random
\twop\ states, $\avg{v} = {\cal O}(1)$, $\avg{y}\sim 1/L^2$, $\br |z|^2 \kt \sim 4/L^3$ and $\br |z| \kt^2=(2/\pi) \br
|z|^2 \kt $.
Thus, the negative term in the pre-concurrence $\left(2(|z|-\sqrt{vy})\right)$ 
is typically larger than the positive term, thus resulting in the probability of a positive
concurrence \emph{decreasing} with increasing $L$ as $1/\sqrt{L}$
 (see ~\cite{Vikram11} for calculations).

However, for \twop\ states promoted from \onep\ states obeying $\sum_i
a_i = 0 $, it is straightforward to show using
$a_{ij}=(a_i+a_j)/\sqrt{L-2}$ that, up to the leading order, \beq
\centering \quad y \approx (a_1+a_2)^2/L, \quad z \approx (1+L a_1
a_2)/L, \quad v \approx 1. \quad
\label{promoZY}
\eeq Therefore, although $z$ appears as a sum of order $L$ number of terms in Eq.~\ref{eq:vyz}, it simplifies for promoted states to this simple form, which implies that both $|z|$ and $\sqrt{y}$ are of the same order of magnitude, namely $1/L$. This follows since $a_i \sim 1/\sqrt{L}$. As $v ={\cal O}(1)$, the concurrence (which is proportional to $|z|-\sqrt{vy}$) in promoted \twop\ states is always in a fine balance between the two competing terms $|z|$ and $\sqrt{y}$. In contrast, for generic \twop\ states, 
the order of $\sqrt{y}$ is $1/L$ which is much larger than the order of $|z|$ which is $1/L^{3/2}$, resulting in the probability of nonzero concurrence scaling as $1/\sqrt{L}$ \cite{Vikram11}.

The probability of finding any two spins entangled when the system is
in a promoted \twop\ state, i.e., $P(C>0)$ is now estimated. This is
approximately same as $P(z^2>y)$, since $v \approx 1$ (Eq.~\ref{promoZY}). Introducing the variables $x_i = \sqrt{L}a_i$, and treating the two $x_i$ to be
independent~\footnote{Strictly speaking, $x_1$ and $x_2$ cannot be independent since $\sum_k a_k = 1$, but the dependence between $a_1$ and $a_2$ is weak,
allowing us to consider $x_1$ and $x_2$ to be independent with a small error}
%(the constraints are distributed spread across all the variables $a_i$, so this is justifiable for large $L$) 
and identically distributed random variables drawn from the standard normal
distribution ${\cal N}(0,1)$, we have \beq
\begin{split}
P(C>0) \approx P(z^2-y >0) = P[ (1-x_1^2)(1-x_2^2)>0]\\ = \mbox{erf}^2 \left(\f{1}{\sqrt{2}} \right)+\mbox{erfc}^2 \left(\f{1}{\sqrt{2}} \right) \approx 0.566,
\end{split}
\label{eq:probability}
\eeq where $\mbox{erfc}(z) \equiv 1-\mbox{erf}(z)$ is the
complementary error function. Thus, $P(C>0)$ is a constant for random promoted \twop\ states and does not decrease with
$L$ as it does for generic \twop\ states. Note that as $v<1$ in reality, the above
can be expected to underestimate the actual probability.  It is also
worth recounting that random \onep\ states have a probability 1 that
the concurrence is nonzero.

The average concurrence of random promoted \twop\ states may also be
estimated by weighted integration over all $x_1,x_2$ where $P(C>0)>0$:
%\begin{equation}
\begin{align} 
\br C\kt & \approx \int\limits_{\prod_i(1-x_i^2) >0} 2(|z|-\sqrt{vy})e^{-\frac{(x_1^2+x_2^2)}{2}}\rmd x_1 \rmd x_2 
         & \approx \frac{0.465}{L}\label{eq:CvsL},
%\label{eq:integral}
\end{align} 

where Eq.~\ref{promoZY} and the assumption of independent marginals
has been used. The final result was obtained by setting $v=1$ and
factoring out the $L$ dependence, and the $L$-independent integral was
evaluated numerically to obtain $0.465$.  The $1/L$ behaviour is to be
compared with generic \twop\ states that have an expectation value of
concurrence $\sim 1/L^2$ \cite{Vikram11}, and that for generic random
\onep\ states which goes as $\sim 1/L$ \cite{ArulSub}.  The promoted
states have, on average, smaller entanglement than \onep\ states, however they are
much larger than what may be expected for generic \twop\ states.

Interestingly, for the promoted \onep\ state, i.e. the all-one state in
the \onep\ sector, the concurrence between any two spins is
$2/L$ and in the \twop\ sector, the all-one state has  $C = \frac{2}{\binom{L}{2}} \left( L-2 -\sqrt{\frac{(L^2-5L+6)}{2}}\right)$, 
which also scales as $1/L$. This deserves a special mention since Eq.~\ref{promoZY} does not hold for the all-one state, yet the
scaling behavior is identical. Thus, on average, the promoted random \twop\ states retain the larger entanglement present in the generic \onep\ states, while at the same time, they are as delocalised as generic \twop\ states.  
Our results are summarised and compared against generic \onep\ and \twop\
states in Table ~\ref{tab:summary_random}.
\begin{table}
\begin{center}
\begin{tabular}{|c|c|c|c|}
\hline
States & $P(C>0)$ & \avg{C} & \avg{\text{IPR}} \\
\hline
One-particle\ & 1 & ${4}/{\pi L}$ & ${3}/{L} $\\
Two-particle\ & ${2\sqrt{2}}/{\sqrt{\pi L}}$ & ${16}/{\pi^{3/2}L^2} $ & ${6}/{L^2}$ \\
 \begin{tabular}{@{}c@{}} Promoted \\ two-particle\ \end{tabular} & $0.566$ & $0.465/L$ & $6/L^2$ \\
\hline
\end{tabular}
\end{center}
\caption{A table comparing the \emph{random} promoted \twop\ states with \emph{random} \onep\ and \twop\ states. 
For the promoted states, $P(C>0)$ is constant and \avg{C} scales as $1/L$ , similar to those of \onep\ states while the localisation
scaling is similar to that of \twop\ states.}
\label{tab:summary_random}
\end{table}
%\section{Promoted states in the spin glass}
\begin{figure}
 \centering
 \includegraphics[width=\columnwidth]{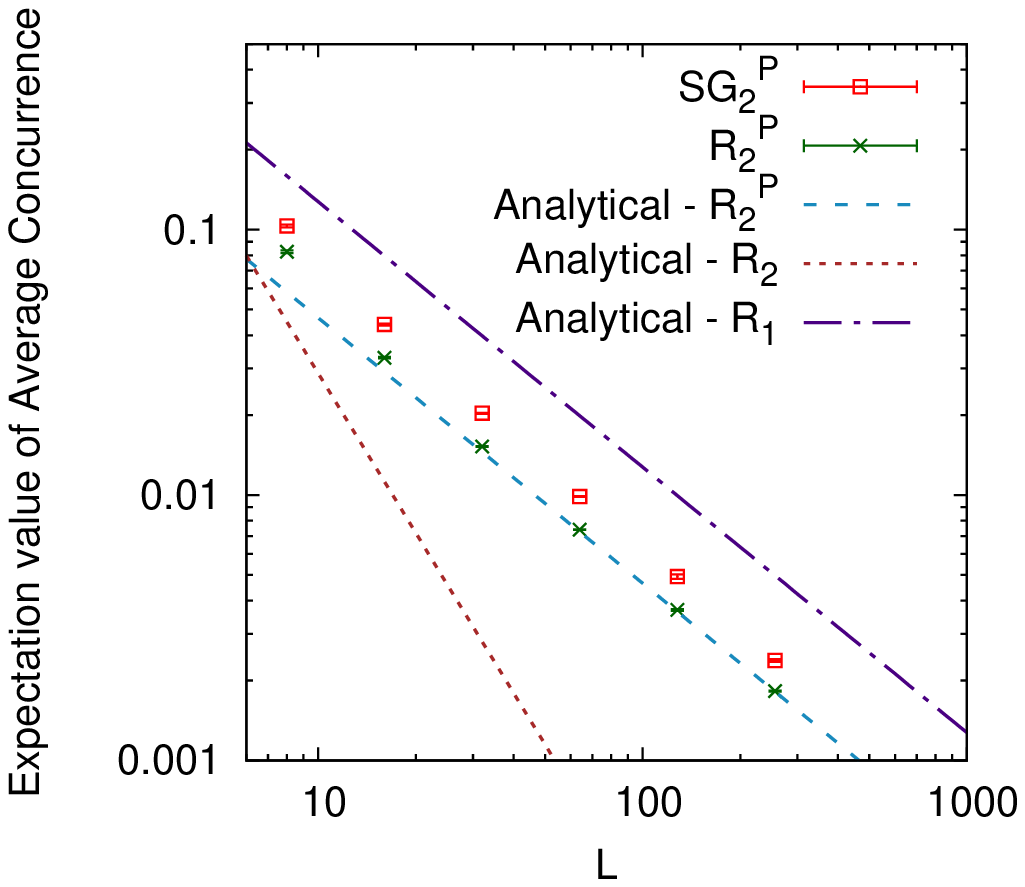}
 \caption{Plot of expectation value of average concurrence showing the scaling behavior with the size of the system, $L$. The meaning of the various labels is the same as from the previous figure, while in addition $R_1$ refers to random \onep\ states. For $SG_2^p$ and $R_2^p$ states, error-bars are included by considering an average over many states (obtained from considering many samples of disorder for small $L$). The error-bars are too tiny to be perceptible.}
 \label{fig:entanglement_scaling}
\end{figure}

  One-particle eigenstates of the \sg\ Hamiltonian in
  Eq.~\ref{eq:ham} (considering a large number of realisations of the
  $J_{ij}$ for various values of $L$) were obtained by exact
  diagonalisation. These were then promoted to \twop\ eigenstates. To
  compare properties, an equal number of \onep\ random states were
  generated to obtain \twop\ promoted random states.
  Fig.~\ref{fig:entanglement_scaling} shows the expectation value of
  the concurrence for the promoted \twop\ eigenstates of the
  \sg\ Hamiltonian and promoted random \twop\ states ($SG_2^p$ and
  $R_2^p$). It is seen that $R_2^p$ states mimic the $SG_2^p$
  eigenstates rather well, and that they behave differently from the
  random \twop\ states ($R_2$)~\cite{Vikram11}.

   Thus using two reasonable assumptions, that $\sum_k a_k = 0$ and 
   that $a_i$ and $a_j$ ($i\ne j$) are marginally independent,
   we have obtained analytical results for random promoted
   \twop\ states. While for large $L$ the agreement with the analysis
   leading to Eq.~\ref{eq:CvsL} with simulated random states gets
   better, the same cannot be said for the \sg\ eigenstates. This
   result is not surprising because the spin glass eigenstates carry
   special structure, which would make them not uniformly distributed
   on the unit sphere in Hilbert space, while random states are
   uniformly distributed, by construction. It is also striking that
   the scaling for generic two particle random states (that goes as
   $1/L^2$ and is also shown for comparison) is indeed very different.
   Thus, the statistical analysis of random \twop\ promoted states
   sheds light on the enhanced entanglement observed in certain
   classes of \sg\ eigenstates.

\section{Summary and future directions}
A central finding of this paper is that when a many-body quantum system is governed by a random Heisenberg Hamiltonian with long range coupling, a special class of eigenstates emerge that are characterised by enhanced entanglement. 
These special eigenstates, shown to be ``promoted-eigenstates", display significantly higher average concurrence compared with the rest of the eigenstates.  

As a first step to understand the peculiarities of  such promoted-eigenstates, the properties of random promoted states were studied by a statistical approach.  It has been proved analytically and confirmed numerically in this work that random \onep\ states, for which the average two-spin entanglement scales as $1/L$, when promoted to have particle-number $2$, shows an average two-spin entanglement that is
lower but {\it still scales} as $1/L$.  This is to be contrasted with the
scaling behaviour of $1/L^2$ for random \twop\ states~\cite{Vikram11}.
In contrast, the localisation of a promoted \twop\ state, as measured by the inverse participation ratio (IPR), is comparable to that of a typical two-particle state.

Thus, our results provide a small but an important step towards understanding how entanglement is shared across small subsystems in a larger system and providing hints on what kind of states should the quantum system be prepared in, to maximise or minimise entanglement as the application might require. From the point of view of random states, this work has introduced the study of promoted random states that will be found in systems with full rotational symmetry.

Two interesting questions seem natural to pursue further:
i) While the random \onep\ states and \onep\ eigenstates have very different distribution
in the concurrence-PR plot (refer Fig.~\ref{fig1}), the distinction is largely reduced after the promotion. It remains to investigate if random promoted states are a good approximation to random
promoted states for all models of the Heisenberg Hamiltonian and at all particle-numbers.
 ii) One can see that the \zerop\ state when promoted to have a particle-number of 1 or 2 (all-one states), the concurrence still scales as $1/L$.
It seems likely that the average two-spin entanglement of promoted \onep\ eigenstates
will continue to scale as $1/L$ in higher particle sectors as well and that of promoted
\twop\ eigenstates will continue to scale as $1/L^2$. This implies that one
can have half-filled states with an average concurrence much higher than what one would normally expect, something that needs further work for verification.

\acknowledgments
The authors thank Dr. V. Subrahmanyam of IITK for a crucial discussion. 
AS acknowledges support from the DST-INSPIRE Faculty Award [DST/INSPIRE/04/2014/002461].
%The Hamiltonian matrices are diagonalized using \texttt{Eigen} package~\cite{eigenweb}. 
%Curve fitting to find optimal parameters were done using SciPy's \texttt{curve\_fit} routine.

\bibliography{ref2014}

\begin{thebibliography}{10}
\expandafter\ifx\csname url\endcsname\relax\def\url#1{\texttt{#1}}\fi

\bibitem{Sougato07}
\Name{Bose S.} \REVIEW{Contemporary Physics}{48}{2007}{13}.

\bibitem{ArulSub}
\Name{Lakshminarayan A. \and Subrahmanyam V.} \REVIEW{Phys. Rev.
  A}{67}{2003}{052304}.
\newline\url{http://link.aps.org/doi/10.1103/PhysRevA.67.052304}

\bibitem{EntLoc1}
\Name{Wang X., Li H. \and Hu B.} \REVIEW{Phys. Rev. A}{69}{2004}{054303}.
\newline\url{http://link.aps.org/doi/10.1103/PhysRevA.69.054303}

\bibitem{EntLoc2}
\Name{Li H., Wang X. \and Hu B.} \REVIEW{J. Phys. A: Math.
  Gen.}{37}{2004}{10665}.
\newline\url{http://iopscience.iop.org/0305-4470/37/44/014}

\bibitem{PhysRevE.82.031130}
\Name{Santos L.~F. \and Rigol M.} \REVIEW{Phys. Rev. E}{82}{2010}{031130}.
\newline\url{http://link.aps.org/doi/10.1103/PhysRevE.82.031130}

\bibitem{PhysRevE.81.036206}
\Name{Santos L.~F. \and Rigol M.} \REVIEW{Phys. Rev. E}{81}{2010}{036206}.
\newline\url{http://link.aps.org/doi/10.1103/PhysRevE.81.036206}

\bibitem{PhysRevA.82.011604}
\Name{Rigol M. \and Santos L.~F.} \REVIEW{Phys. Rev. A}{82}{2010}{011604}.
\newline\url{http://link.aps.org/doi/10.1103/PhysRevA.82.011604}

\bibitem{PhysRevE.87.012125}
\Name{Ikeda T.~N., Watanabe Y. \and Ueda M.} \REVIEW{Phys. Rev.
  E}{87}{2013}{012125}.
\newline\url{http://link.aps.org/doi/10.1103/PhysRevE.87.012125}

\bibitem{Wooters}
\Name{Hill S. \and Wootters W.~K.} \REVIEW{Phys. Rev. Lett.}{78}{1997}{5022}.

\bibitem{Woottersentform}
\Name{Wootters W.~K.} \REVIEW{Phys. Rev. Lett.}{80}{1998}{2245}.
\newline\url{http://link.aps.org/doi/10.1103/PhysRevLett.80.2245}

\bibitem{Karthik07}
\Name{Karthik J., Sharma A. \and Lakshminarayan A.} \REVIEW{Phys. Rev.
  A}{75}{2007}{022304}.

\bibitem{Brown:08}
\Name{Brown W.~G., Santos L.~F., Starling D.~J. \and Viola L.} \REVIEW{Phys.
  Rev. E}{77}{2008}{021106}.

\bibitem{dukesz2009interplay}
\Name{Dukesz F., Zilbergerts M. \and Santos L.~F.} \REVIEW{New Journal of
  Physics}{11}{2009}{043026}.

\bibitem{beugeling:2015}
\Name{Beugeling W., Andreanov A. \and Haque M.} \REVIEW{J. Stat.
  Mech}{2015}{2015}{P02002}.

\bibitem{viola2007generalized}
\Name{Viola L. \and Brown W.~G.} \REVIEW{Journal of Physics A: Mathematical and
  Theoretical}{40}{2007}{8109}.

\bibitem{giraud2007entanglement}
\Name{Giraud O., Martin J. \and Georgeot B.} \REVIEW{Physical Review
  A}{76}{2007}{042333}.

\bibitem{Vikram11}
\Name{Vijayaraghavan V.~S., T.Bhosale U. \and Lakshminarayan A.} \REVIEW{Phys.
  Rev. A}{84}{2011}{032306}.

\bibitem{Santos2004}
\Name{Santos L.~F., Rigolin G. \and Escobar C.~O.} \REVIEW{Phys. Rev.
  A}{69}{2004}{042304}.
\newline\url{http://link.aps.org/doi/10.1103/PhysRevA.69.042304}

\bibitem{Carlos2005}
\Name{Mej\'{\i}a-Monasterio C., Benenti G., Carlo G.~G. \and Casati G.}
  \REVIEW{Phys. Rev. A}{71}{2005}{062324}.
\newline\url{http://link.aps.org/doi/10.1103/PhysRevA.71.062324}

\bibitem{Bera2016}
\Name{Bera S. \and Lakshminarayan A.} \REVIEW{Phys. Rev. B}{93}{2016}{134204}.
\newline\url{http://link.aps.org/doi/10.1103/PhysRevB.93.134204}

\bibitem{kendon2002bounds}
\Name{Kendon V.~M., \ifmmode~\dot{Z}\else \.{Z}\fi{}yczkowski K. \and Munro
  W.~J.} \REVIEW{Phys. Rev. A}{66}{2002}{062310}.
\newline\url{http://link.aps.org/doi/10.1103/PhysRevA.66.062310}

\bibitem{kendon2002typical}
\Name{Kendon V.~M., Nemoto K. \and Munro W.~J.} \REVIEW{Journal of Modern
  Optics}{49}{2002}{1709}.

\bibitem{scott2003entangling}
\Name{Scott A.~J. \and Caves C.~M.} \REVIEW{Journal of Physics A: Mathematical
  and General}{36}{2003}{9553}.

\bibitem{Bhosale2012}
\Name{Bhosale U.~T., Tomsovic S. \and Lakshminarayan A.} \REVIEW{Phys. Rev.
  A}{85}{2012}{062331}.
\newline\url{http://link.aps.org/doi/10.1103/PhysRevA.85.062331}

\bibitem{Amico08}
\Name{Amico L., R.~Fazio A.~O. \and Vedral V.} \REVIEW{Rev. Mod.
  Phys.}{80}{2008}{517}.

\bibitem{Connor2001}
\Name{O'Connor K.~M. \and Wootters W.~K.} \REVIEW{Phys. Rev.
  A}{63}{2001}{052302}.
\newline\url{http://link.aps.org/doi/10.1103/PhysRevA.63.052302}

\bibitem{WoottersRandom}
\Name{Wootters W.~K.} \REVIEW{Foundations of Physics}{20}{1990}{11}.

\bibitem{brody81}
\Name{Brody T.~A., Flores J., French J.~B., Mello P.~A., Pandey A. \and Wong
  S.~S.~M.} \REVIEW{Rev. Mod. Phys.}{53}{1981}{385}.

\end{thebibliography}
\bibliographystyle{eplbib}

\end{document}